\RequirePackage{fix-cm}
\documentclass[smallextended]{svjour3}       
\smartqed  
\usepackage{graphicx}
\usepackage{import}
\usepackage{amsmath,amssymb,amsfonts}
\usepackage{algorithmic}
\usepackage{graphicx}
\usepackage{textcomp}
\usepackage{xcolor}
\usepackage{svg}

\usepackage{tabularx}

\usepackage{mdwlist}
\usepackage[ruled,vlined,shortend, linesnumbered]{algorithm2e} 
\usepackage[]{algorithmic} 
\usepackage{booktabs}
\usepackage{color}

\usepackage{amsthm,amssymb}
\usepackage{amsmath}
\usepackage{amsfonts}
\usepackage{bm} 
\usepackage{multirow}
\usepackage{mathrsfs}
\usepackage{amsbsy}
\usepackage[mathscr]{eucal} 
\usepackage[flushleft]{threeparttable}
\usepackage[hidelinks]{hyperref}
\usepackage{comment}
\usepackage{paralist}
\usepackage{pifont}

\usepackage{bbding} 

\def\BibTeX{{\rm B\kern-.05em{\sc i\kern-.025em b}\kern-.08em
    T\kern-.1667em\lower.7ex\hbox{E}\kern-.125emX}}

\usepackage[colorinlistoftodos]{todonotes}
\usepackage{xcolor}

\theoremstyle{definition}

\usepackage{amsthm}
\usepackage{amsmath}
\usepackage{amssymb}
\usepackage{thmtools}
\usepackage{thm-restate}
\usepackage{hyperref}

\usepackage{mathtools}
\usepackage[
backend=bibtex,
style=ieee,
sorting=none
]{biblatex}
\DeclareUnicodeCharacter{2212}{-}

\begin{document}

\title{Large-scale Grid Optimization: The Workhorse of Future Grid Computations
\thanks{}
}


\author{Amritanshu Pandey and \\
        Mads R. Almassalkhi and \\
        Samuel Chevalier 
}


\institute{A. Pandey \at
              University of Vermont \\
              \email{amritanshu.pandey@uvm.edu}  
            \and
            M. Almassalkhi \at
              University of Vermont \\
              \email{malmassa@uvm.edu}           
           \and
           S. Chevalier \at
            University of Vermont \\
              \email{samuel.chevalier@uvm.edu}           
}

\date{Published: 10 July 2023}

\maketitle

\begin{abstract}

\textbf{Purpose:} The computation methods for modeling, controlling and optimizing the transforming grid are evolving rapidly. We review and systemize knowledge for a special class of computation methods that solve large-scale power grid optimization problems. \textbf{Summary:} Large-scale grid optimizations are pertinent for, amongst other things, hedging against risk due to resource stochasticity, evaluating aggregated DERs' impact on grid operation and design, and improving the overall efficiency of grid operation in terms of cost, reliability, and carbon footprint. We attribute the continual growth in scale and complexity of grid optimizations to a large influx of new spatial and temporal features in both transmission (T) and distribution (D) networks. Therefore, to systemize knowledge in the field, we discuss the recent advancements in T and D systems from the viewpoint of mechanistic physics-based and emerging data-driven methods.  \textbf{Findings:} We find that while mechanistic physics-based methods are leading the science in solving large-scale grid optimizations, data-driven techniques, especially physics constrained ones, are emerging as an alternative to solve otherwise intractable problems. We also find observable gaps in the field and ascertain these gaps from the paper's literature review and by collecting and synthesizing feedback from industry experts.

\end{abstract}

\section{Introduction}
Modern power grids rely on optimization computations for reliable, resilient, and secure operation and design. These computations dictate many grid mechanisms, such as setting nodal electricity prices (e.g., \cite{scuc_hippo}, \cite{scuc_warmstart}, \cite{sto_scuc_ISO}), building new capacity (e.g., \cite{tep_benders}, \cite{tep_canizes}, \cite{tep_bilevel}), decarbonizing grid operation (e.g., \cite{ce_nrel_india}, \cite{ce_newman_guide}, \cite{ce_reeds_review_open_source_tools}), and estimating the operational grid state \cite{schweppe1974static} \cite{abur2004power}. There is a wide spectrum of optimization techniques for these applications, ranging from stochastic (e.g., \cite{sto_scuc_2018}, \cite{sto_scuc_ISO}, \cite{sto_scuc_minmax_robust}) to deterministic methods with both continuous (e.g., \cite{mpopf_jabr}, \cite{mpopf_global}) and integer variables (e.g., \cite{scuc_hippo}, \cite{sto_scuc_ISO}).

Until recently, except for a few (e.g., \cite{scuc_hippo}, \cite{scopf_go_survey}), most optimization problems for power grid applications were small-scale ($\leq$ 100k variables) and realizable on a single compute resource. However, the condition no longer holds due to rapid changes to the modern electric grid. The scale and complexity of grid optimizations are increasing due to the exponential increase in \textit{spatial} and \textit{temporal} grid features. This phenomenon is further exacerbated as many of these emerging features are stochastic. Therefore, in more recent grid optimization literature, we observe methods that combine and include complexities due to both spatial and temporal features (for instance, \cite{scopf_mp_lookahead} considers both look-ahead time intervals and a higher number of security constraints within the optimal power flow (OPF) problem). Similarly, we observe significant growth in stochastic methods for grid analyses, including those based on data-driven techniques (e.g., \cite{cc_data_driven_gp}).

We find the growth in spatial features is mainly due to the large-scale penetration of distributed energy resources (DERs) in the distribution grid (D). In the past, the transmission (T) and distribution grids could operate independently, the power constantly flowed from T to D, and the communication between various T\&D entities was minimal and sporadic. Hence, an independent analysis of T and D entities was sufficient. However, this approach will not suffice in the near future, and we will have to optimize T\&D resources together, resulting in many more spatial features. Other factors also contribute to growth in spatial features, such as the need to include a higher number of security constraints within the optimization problem definition. 

The growth in the temporal features is mainly due to growing uncertainty in renewable generation output and rapid energy storage systems (ESS) installations. Making decisions in operations or planning with growing uncertainty and more storage requires looking ahead into the time horizon. Therefore, instead of solving single-period simulation and optimization problems, the grid planners and operators will have to solve far larger multi-period problems in the future. Additionally, forecasting and planning for the net-zero future electric grid scenarios also necessitate large multi-period optimizations. For instance, a recent study of renewable penetration in India looked decades into the future and used an elaborate capacity expansion optimization model \cite{ce_nrel_india}.

Another source fueling the growth in the scale of grid optimization is the high uncertainty associated with emerging grid resources. Stochastic optimizations are better suited to analyzing systems with these resources and many stochastic optimization techniques are inherently large-scale optimizations with the underlying curse of dimensionality. Observing these emerging patterns in grid optimizations, in the subsequent sections of this paper, we discuss and systemize relevant literature in the field of large-scale grid optimization for transmission networks (see Section \ref{sec:transmission}), distribution networks (see Section \ref{sec:distributed}), and the combination of the two (see Section \ref{sec:combined}). For these categories, we will systemize knowledge from the viewpoint of both mechanistic physics-based as well as data-driven techniques. While physics-based techniques make up the majority of the recent literature on large-scale grid optimization, emerging data-driven techniques have shown potential in tackling problems otherwise intractable. We will conclude (in Section \ref{sec:gaps}) by documenting observable gaps in the field of large-scale grid optimization by synthesizing knowledge from discussions with industry experts and the literature review in subsequent sections.

\section{Advancements in Large-Scale Optimization: A Taxonomy}

\begin{figure*}[!ht]
    \centering\includegraphics[width=0.95\textwidth]
    {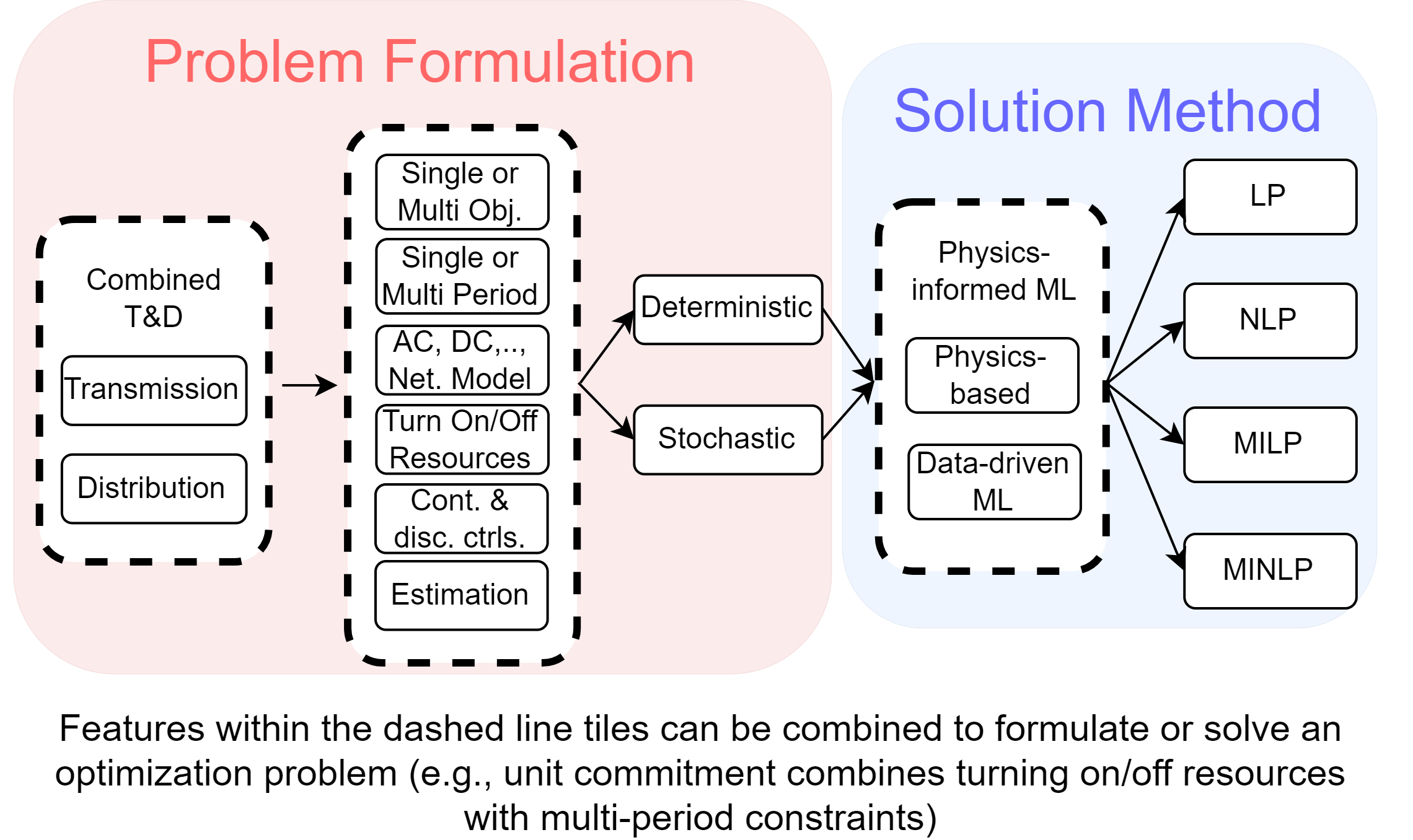}
    \caption{Taxonomy for large-scale grid optimizations. The left of the figure describes various facets of grid optimization problem formulation. The right of the figure describes the relevant solution techniques.}
    \label{fig:taxonomy}
\end{figure*}

From the viewpoint of real-world applications, large-scale optimizations emerge when addressing specific problems in high-voltage transmission grids, low-voltage distribution grids, and, more recently, when the two are combined. Optimizations for other smaller-sized networks generally do not require large-scale computing methods. In Figure \ref{fig:taxonomy}, we sketch a non-exhaustive taxonomy for large-scale grid optimizations. In Table \ref{tab:application_table}, building on the taxonomy, we enumerate several specific instances of large-scale grid optimizations. In the following sections, we discuss the state-of-the-art in some critical instances of large-scale optimizations. Note that the optimization instances (in Sections \ref{sec:transmission} through \ref{sec:combined}) and the corresponding citations in this paper are non-exhaustive, as the goal solely is to synthesize \textit{recent} advancements in large-scale optimizations pertinent to grid planners, policy researchers, and operators. 

\begin{table*}[htbp]
\caption{Instantiating large-scale grid optimization by non-exhaustive taxonomy in Fig. \ref{fig:taxonomy}.}
\label{tab:application_table}
\begin{tabular}
{>{\raggedright}p{0.22\linewidth}|>{\raggedright}p{0.12\linewidth}|p{0.08\linewidth}|>{\raggedright}p{0.21\linewidth}|p{0.07\linewidth}|p{0.08\linewidth}}
\hline\noalign{\smallskip}
\textbf{Application} & \textbf{E.g. literature} & \textbf{Net. Type$^2$} & \textbf{Features} & \textbf{Stoch. Opt?$^3$} & \textbf{Problem Type} \\ 
\noalign{\smallskip}\hline\noalign{\smallskip}
Capacity Expansion Problem (CEP)
& \cite{ce_nrel_india}, \cite{ce_scholz}, \cite{ce_splicing}, \cite{ce_reeds_review_open_source_tools},
\cite{ce_NEMS_2019}
& T 
& Copper plane or simple network model, Multi-period, 
& N
& MILP \\
\hline
Production Cost Model (PCM)
& \cite{pcm_ce_linking}, \cite{pcm_multi}, \cite{pcm_powergem}, \cite{pcm_riia_miso}, \cite{pcm_switch2}
& T
& Simple network model (Linearized), Multi-period, 
& N
& MILP \\
\hline
Security Constrained Unit-commitment (SCUC)
& \cite{scuc_challenges}, \cite{scuc_hippo}, 
\cite{scuc_ml_1}, \cite{scuc_neighborhood_search}, \cite{scuc_warmstart}
& T
& Simple network model (Linearized), Multi-period, 
& N
& MILP \\
\hline
Stochastic Unit-commitment (Sto. SCUC)
& \cite{sto_scuc_2018}, \cite{sto_scuc_chance_perform2}, \cite{sto_scuc_dvorkin}, \cite{sto_scuc_ISO}, \cite{sto_scuc_2018}
& T
& Simple network model (Linearized), Multi-period, 
& Y
& MILP \\
\hline
Multi-period OPF (MPOPF)
& \cite{mpopf_mads}, 
\cite{mpopf_aayushya},
\cite{mpopf_parallel_1},
\cite{mpopf_jabr},
\cite{mpopf_olaf}
& T, D
& AC or relaxed network model, Multi-period, 
& Y
& NLP \\
\hline
Security-constrained ACOPF (SCOPF)
& \cite{go_bienstock}, 
\cite{go_jereminov}, \cite{go_mcnamara}, 
\cite{go_coffrin}, \cite{Crozier:2022_pmaps}
& T 
& AC network model, Single period, 
& N
& NLP \\
\hline
Stochastic ACOPF (Sto. ACOPF)
& 
\cite{Agarwal:2022}, \cite{sto_opf_mc_data},
\cite{sto_survey_roald}, \cite{chance_constrained_bienstock},
\cite{sto_scuc_dvorkin}
& T 
& AC network model, Single period, 
& Y
& NLP \\
\hline
Transmission Expansion Problem (TEP)
& \cite{tep_benders}, 
\cite{tep_bilevel}, 
\cite{tep_canizes},  
\cite{tep_conejo}, 
\cite{tep_genetic_algo}
& T
& AC network model, Multi-period, 
& N
& MINLP \\
\hline
Combined T\&D AC Optimization (TD-ACOPF)
& \cite{TD_opti1}, 
\cite{TD_opti_control},
\cite{TD_opti_national_labs},
\cite{se_dist_global},
\cite{TD_SE}
& T\&D
& AC constraints, single-period
& N
& NLP \\
\hline
AC State Estimation (ACSE)
&\cite{Chevalier:2021,zamzam2019data}, \cite{se_dist_xie}, 
\cite{se_dist_guo}, 
\cite{se_dist_minot}, 
& T, D
& AC constraints, single-period
& Y
& NLP or LP \\
\hline
Distribution Optimal Power Flow (DOPF)
& \cite{Manish:2020}, \cite{paudyal_optimal_2011}, \cite{bernstein_load_2018}, \cite{stai_solution_2021}, \cite{dallanese_optimal_2018}
& D
& AC or linearized constraints, single-period
& N
& NLP or LP\\
\hline
Stochastic Distribution Optimal Power Flow (Sto. DOPF)
& 
\cite{dallanese_chance-constrained_2017}, \cite{almassalkhi_hierarchical_2020}, 
\cite{sto_DOPF1}, 
\cite{sto_data_driven},
\cite{sto_DOPF_nali}
& D
& AC or linearized constraints, single-period
& Y
& NLP or LP\\
\noalign{\smallskip}\hline
\end{tabular}
\\
\footnotesize{
1. This is a non-exhaustive list and is only meant to enumerate how the various large-scale grid optimizations can be categorized based on the taxonomy in Figure \ref{fig:taxonomy}.\\
2. T refers to transmission, D refers to distribution, T\&D refers to combined T\&D}\\
3. Stoch. Opt. indicates whether or not stochastic optimization techniques are used to solve this problem.
\end{table*}

\section{Optimizations for High-Voltage Transmission Grids} \label{sec:transmission}

Several computations solved by transmission grid operators, planners, and policymakers are framed as optimizations. Many of these are large-scale or have significant potential to become one due to emerging patterns in the power grid. We find industry experts concur \cite{isone_private}, \cite{kwok_private}. In this section, we focus on key transmission grid optimizations from mechanistic physics-based and data-driven viewpoints, discussing the emerging advancements in the last half-dozen years.

\subsection{Mechanistic Physics-based Solution Techniques}

Mechanistic physics-based solution techniques are the bedrock of industry-run large-scale optimizations in transmission networks. They are seeing a fast evolution as the needs of transmission network operators and planners emerge. We focus on a critical few instances of these advancements. We begin with applications that optimize the network over the longest time horizon and gradually narrow it down to those that only optimize it over a single time snapshot. We also cover stochastic instances due to their growing need by grid operators and planners \cite{isone_private}.

\textit{Capacity Expansion Problem}: The large-scale grid problems that have the longest horizon today are the capacity expansion problems (CEP) solved on \textit{energy system models}. These simulate transmission and generation investment in capacity upgrades, given inputs and assumptions about future load demand, fuel prices, policy, regulation and cost and performance of various technologies. These problems can be run over a horizon of 10 years or more. They answer questions like: what mix of generators should be built for net-zero goals while satisfying future electricity demand? They use a copper plate model or a simplistic network model with linear constraints. They are formulated as large-scale linear programs (LPs) and solved using simplex or interior point algorithms with tens of millions of decision variables \cite{ce_reeds_review_open_source_tools}. Some of the prevalent CEP models are the Regional Energy Deployment System  (ReEDs) \cite{ce_reeds_2020}, National
Energy Modeling System (NEMS) \cite{ce_NEMS_2019}, MARKet ALlocation (MARKAL)\cite{ce_MARKAL}, and Integrated Planning Model (IPM). 

We find that \cite{ce_reeds_review_open_source_tools} is a good resource for learning about the most recent advancements in CEP, and we highlight the key ones here. A recent study by \cite{ce_scholz} found that interior point methods, in general, outperform primal and dual simplex methods for large-scale CEPs. A recent NREL study \cite{ce_reeds_review_open_source_tools} also supports that interior point methods are faster than simplex methods for CEPs. Another resource that compares various linear solvers for CEP can be found in \cite{ce_newman_guide}.

Other than LP solution techniques, recent advancements in CEP have stemmed from changes to the problem formulation. Most advances are based on model reduction or decomposition techniques \cite{ce_scholz}. Model reductions include techniques such as slicing and spatial and temporal aggregation. Model reduction focuses on reducing the problem size to achieve numerical efficiency, whereas model decomposition uses parallel techniques to solve the exact problem via problem decomposition. A recent study by NREL \cite{ce_nrel_india} achieves model reduction through time-slicing. This study uses only 35-time slices to approximate India's yearly energy behavior. Alternatively, one could solve a CEP model with more time slices 
by applying decomposition techniques such as Dantzig-Wolfe, Lagrangian, and Benders decomposition. These have been discussed in \cite{ce_scholz}.


\textit{Production Cost Modeling:} Closer in scope to CEP are the production-cost modeling problems (PCM). These study the network over a much shorter time horizon but model the grid at a higher fidelity. They minimize generation costs over a horizon and obey reliability requirements, and they include detailed load, transmission, and generation fleet data. PCMs are extensively used for operation economics and portfolio management by forecasting future nodal pricing. Commercial vendors have made the most recent advancements in PCM. These include Enerlytix's Power System Optimizer, NREL's Resource Planning Model (RPM), GE's MAPS, and PLEXOS and Aurora by Energy Exemplar. In the open literature, there are also advancements. \cite{pcm_multi} recently proposed a multi-operator PCM targeting the sub-optimalities due to separate PCMs between operators. An open-source tool, Switch 2.0 \cite{pcm_switch2}, has shown that it can achieve comparable results with GE-MAPS for PCM models for the Hawaiin network. There is also a push to make PCM more consistent with day-ahead SCUC markets to improve the accuracy of their economic forecasts \cite{pcm_powergem}. This will require higher computation power and higher-granularity grid models. But more importantly, it will require fine-tuning the PCM models for each grid operator to include their market rules. There is also literature on linking the PCM models with higher fidelity ACPF and ACOPF models  \cite{pcm_riia_miso} and longer horizon CEP models \cite{pcm_ce_linking}. The idea is that output from the PCM model should be consistent with AC steady-state 
 and CEP models. This is not the case today and generally requires specialized optimization capabilities (see the use of commercial circuit-based optimization \cite{ecp}, \cite{circuit_pandey} for MISO's RIIA study \cite{pcm_riia_miso}).


\textit{Security-constrained Unit Commitment:} A security-constrained unit commitment (SCUC) is similar to PCM but is for a shorter time-horizon and includes other critical features, which PCM does not consider. It schedules the output of power-generating units while accounting for the transmission security constraints. In the U.S., independent system operators (ISOs) are primarily the entities that run SCUC to commit generators and set prices in day-ahead (DA) and real-time markets. We will focus on DA SCUC as it is a more challenging optimization problem, and we find there are calls for improvements based on discussions with industry experts \cite{isone_private}, \cite{kwok_private} (see, also Section \ref{sec:gaps}).

To address these needs, many improvements are being researched and proposed. Recent works have proposed to improve SCUC runtime by reducing the number of binary variables and security constraints \cite{scuc_neighborhood_search}. Then, there are works that propose improved modeling of grid resources (for instance, see \cite{scuc_cc} for modeling combined-cycle configuration transitioning). Similarly, a large collaborative project High-Performance Power-Grid Optimization (HIPPO), between many entities (including an ISO and a national laboratory), has proposed advancements in the SCUC algorithm to improve scale, performance, and efficiency \cite{scuc_hippo}, mostly through the use of high-performance computing. Warm start techniques are also being proposed to improve the run-time of the SCUC algorithm in the real world. \cite{scuc_warmstart} shows 30\%-40\% improvements in speed with the use of warm start in MISO footprint. An excellent review of recent advancements in SCUC can be found in \cite{scuc_challenges}. 

\textit{Stochastic SCUC and Markets:} Due to the stochastic nature of emerging market resources, there is significant interest in transitioning to stochastic SCUC in markets. The problem is not new, and many scholastic works from a decade ago exist (e.g., key ones include \cite{sto_scuc_og}, \cite{sto_scuc_og1}). However, these did not make it to practice, and as such, there are continuing advancements due to the importance of the problem \cite{sto_scuc_minmax_robust}, \cite{sto_scuc_ISO}, \cite{sto_scuc_2018}, \cite{sto_scuc_dvorkin}. A recent program, 
Performance-based Energy Resource Feedback, Optimization, and Risk Management" by Advanced Research Projects Agency-Energy (ARPA-E), aims to bring together industry and academics to develop  practical solutions to hedge against risk due to stochastic resources in the market. Subsequently, we are beginning to see preliminary results from this program on stochastic treatment of electricity markets (e.g., \cite{sto_scuc_perform1}, \cite{sto_scuc_chance_perform2})


\textit{Security-constrained ACOPF (SCOPF):} In the real world, SCUC is run in tandem with security analysis (also called simultaneous feasibility test, SFT) to ensure the AC feasibility of output dispatch during contingencies. The two can be combined within a Security-constrained ACOPF (SCOPF) framework in an ideal setting. A recent study by the Federal Energy Regulatory Commission (FERC) estimates that such a transition can save billions of dollars in energy costs \cite{ferc_history_of_opf}. Therefore, to advance the state-of-the-art in SCOPF methods, ARPA-E recently launched a grid optimization challenge with large-scale optimizations of sizes of up to 900M continuous variables and 250M discrete variables \cite{scopf_go_size}. The competing researchers in this program have proposed many advancements for large-scale SCOPF. Some of the successful approaches include \cite{scopf_hassan}, \cite{scopf_petra}, \cite{go_sun}, \cite{go_bienstock}, \cite{go_bienstock}, \cite{go_kyri}, and \cite{go_jereminov}.
An excellent survey describing various approaches in the competition can be found in \cite{scopf_go_survey}. In general, we find that most methods used well-established NLP tools such as IPOPT \cite{IPOPT} and Knitro \cite{knitro} as their core engine, and the majority of advancements were made by improving the initial conditions \cite{scopf_hassan}, incorporating efficient parallelism through innovative extensions of ADMM-like algorithms \cite{go_sun_admm_gholami}, and filtering the set of critical contingencies through domain knowledge \cite{go_bienstock}, \cite{go_kyri}. Some approaches \cite{go_jereminov} \cite{go_mcnamara} did not use general commercial NLP solvers and instead developed their domain-focused versions. For example, \cite{go_jereminov} and \cite{go_mcnamara} used a circuit-theoretic-based optimization approach. Besides advancements made during grid optimization competition, there are other emerging efforts at improving large-scale SCOPF performance. Most of these exploit the underlying structure of these problems to improve performance. One such effort is described in \cite{scopf_beltistos} and develops a robust interior point method for large-scale SCOPF. 


\textit{Multi-period ACOPF:} MP-OPF is a class of NLP ACOPF problems that include network physics over multiple time horizons. In contrast to SCUC, these problems do not include integer variables but provide AC-feasible solutions. The need for these analyses is driven primarily by the growing number of energy-constrained batteries, dynamic line ratings, generator ramp limits, and stochastic resources \cite{mpc_mads1}, \cite{mp_mads3}. Many recent solution approaches have been proposed for MP-OPF, ranging from the analysis of DER resources \cite{mpopf_mads} to those that hedge against risk from stochastic resources \cite{mpopf_jabr}, \cite{mpopf_parallel_1}. Many have extended original formulations to develop global optimizers using semi-definite programming for networks with certain characteristics \cite{mpopf_global}, \cite{mpopf_parallel_2}. But, these approaches are applied only to small networks and, subsequently, do not need decomposition techniques to solve the overall problem. By the nature of problem definition, multi-period analysis for larger real-world power grids will be large-scale problems requiring distributed and parallel techniques. More recent techniques such as \cite{mpopf_olaf} and \cite{mpopf_aayushya} have focussed on that facet of the problem. Others have looked at the combination of multi-period and security-constrained ACOPF in look-ahead SCOPF \cite{scopf_mp_lookahead}.

\textit{Other Problems:} Scale and complexity are increasing for other specialized grid optimizations, especially in the planning realm. For instance, grid planners seek robust optimization tools to help with resource and transmission planning beyond the ten-year horizon (e.g., 2040 and 2050) while considering uncertainties \cite{isone_private}. This is an active research area and, in the U.S., requires robust solutions for optimizations on large networks such as the Eastern Interconnect. Generally, for such problems, the runtime is less of a focus than the method's robustness. Large transmission expansion problems (TEPs) represent one such instance of these optimizations and, if solved exactly, are MINLPs that are solved over multiple periods. However, most solution techniques for TEPs today do not solve the exact problem \cite{tep_early_review}. They either relax the nonlinear constraints to solve the MILP problem \cite{tep_benders} or the integrality constraints to solve the NLP problem \cite{tep_genetic_algo}. With tight relaxations, NLP problems can provide epsilon close AC feasible solutions, but they do not provide guarantees on the optimality. MILP-like approaches do not satisfy AC feasibility but can provide better optimality guarantees. Other critical emerging optimizations in planning are those that consider resource uncertainty. TEPs that consider resource uncertainty include \cite{tep_conejo}, \cite{tep_load_uncertainty, tep_uncertainty}. There is also expansive literature on stochastic optimal power flow-based problems (see \cite{sto_survey_roald}). These are for both grid operation and planning. Common solution techniques for stochastic OPF-like problems include stochastic programming methods like Monte-Carlo and scenario analysis (e.g., \cite{sto_opf_mc_data}) and robust optimization techniques like chance-constrained-based methods \cite{chance_constrained_bienstock}. 

Another optimization type that caters to stochastic variables is the estimation techniques for power grid states. These are also growing in complexity and scale with many distributed approaches proposed in recent literature \cite{se_dist_guo}, \cite{se_dist_minot}, \cite{se_dist_xie}, \cite{se_large_scale}, \cite{se_dist_pmu}. The need for distributed algorithms is, in part, due to increasing problem size \cite{se_dist_minot}, increasing interdependency between T\&D networks  \cite{se_dist_global}, and privacy-protecting constraints for estimation of grid states \cite{se_dist_xie}. In general, we find that the complexity and scale of estimation algorithms are likely to increase due to the rapid introduction of numerous grid elements (e.g., DERs, EVs) and the introduction of \textit{unknown} topology and parameters into the mix.

\subsection{Emerging data-driven techniques for transmission networks}
In contrast to mechanistic, physics-based solvers, an abundance of data-driven, Machine Learning (ML), and Artificial Intelligence (AI) solution methodologies are emerging in the literature. While these tools have predominantly flourished within the safe confines of academic and research communities, 80\% of recently polled power-sector executives worry their companies risk going out of business if they do not start to scale AI services by 2025~\cite{AI_insights}. Recent reviews papers~\cite{Donti:2021} and~\cite{Duchesne:2020} holistically document the many ways in which ML is being applied to problems associated with power and energy system operation, while~\cite{van2021machine} and~\cite{Hasan:2020} focus specifically on methods which solve power flow-type optimization problems with ML-based models. Many of these models learn input-output regression mappings, thus attempting to bypass numerical solvers altogether (so-called ``end-to-end" learning). Despite the success of these methods, many of them have not been tested on realistically scaled systems. In the following paragraphs, we focus on the methods which \textit{have} been tested on large-scale systems and show promise for future development. Furthermore, since ML has been applied to an enormous number of transmission-level problems, we concentrate on applications related to (\textit{i}) ACOPF, (\textit{ii}) UC, (\textit{iii}) data-driven constraint screening, (\textit{iv}) operational decision augmentation, and (\textit{v}) using training tools from the ML community to solve conventional optimization problems.

\textit{Learning ACOPF}: For the ACOPF problem, a number of recent works have focused on learning either solutions \cite{Zamzam:2020,gnn-acopf}, warm/hot-start predictions \cite{gnn-acopf-warm,Baker:2019}, or numerical solver iterations \cite{baker2020learning} (i.e., modeling the steps taken by an interior point solver). One of the biggest challenges associated with learning solutions to constrained optimization problems, however, is the enforcement of feasibility. In an attempt to overcome this challenge, the ``DC3" tool \cite{acopf-dnn-donti} learned a partial OPF solution and then projected the missing variables back into the feasible space through an elegant differentiable projection; this approach, however, was never scaled.

For more direct constraint enforcement, many works have utilized principles from Lagrange duality. Authors in~\cite{Nellikkath:2022} used a physics-informed architecture by aggressively regularizing with KKT condition discrepancy, where the NN itself predicted the dual variables. Authors in~\cite{fioretto2020predicting} designed a deep-learning OPF tool called OPF-DNN. Their training methodology exploited Lagrange duality by simultaneously solving for Neural Network (NN) model weights and the sets of Lagrange multipliers which limited constraint violations. The tool showed a high degree of prediction accuracy on test cases with up to 300 buses. OPF-DNN was then improved and expanded in~\cite{Chatzos:2020}, where it was shown to predict solutions to within 0.01\% of optimality on the 3,400-bus French transmission system test case. Using a similar methodology, the same group of authors in~\cite{mak2022learning} proposed a consensus-based algorithm for solving a decentralized ACOPF problem. By learning both the primal and dual consensus variables, the procedure warm-started an ADMM routine by directly predicting primal and dual coupling variable values. The algorithm performed well on large-scale systems with up to 6,700 buses and 320 coupling branches, converging 6 times faster than conventional ADMM. A similar approach was adopted in~\cite{Biagioni:2022}, where a recurrent NN was used to predict consensus variables for large DC-OPF applications. Larger still, ~\cite{park2023compact} used principal component analysis (PCA) to explicitly compress the feature space of massive ACOPF problems. Their tool, Compact Learning, was successfully tested on a PGLib test case with 30,000 buses; according to the authors (in January of 2023), this represents the "largest ACOPF problem(s) to which an end-to-end learning scheme has been applied."

\textit{Learning UC}: Learning-based approaches have also been developed for various UC and SCUC problems \cite{Yang:2021}. Using the MIPLearn package \cite{Alinson:2020}, the authors in \cite{scuc_ml_1} learned generator commitment schedules via support vector machine and k-nearest neighbor (kNN) algorithms; solution predictions accelerated DCUC solutions up to an order of magnitude on systems of up to 6,515 buses. Similarly, \cite{Pineda:2020}, used data-driven constraint screening (learned via kNN) to accelerate DCUC solvers on systems with up to 2000 buses. Within the context of Lagrangian relaxation, \cite{Jianghua:2023} used a NN to predict/warm-start the solutions of various relaxed sub-problems; when the predictions were not of sufficient quality, a branch-and-bound backup procedure corrected the NN prediction and re-tuned the NN in an online fashion. This creative approach showed promising results for accelerating UC results on the 2383-bus Polish system. In order to tackle the ACUC problem, \cite{Kody:2022} used compact NNs to first learn piecewise linear approximations of the AC power flow equations. These approximations were then embedded as power flow constraints inside an ACUC problem.

\textit{Learning inactive constraints}: Increasingly, researchers are using learning-based approaches to screen out potentially inactive inequality constraints\footnote{Every inequality constraint, even if inactive, introduces a new dual variable into an optimization formulation, thus increasing problem complexity for primal-dual solvers.}. This approach is highly desirable because solutions to the smaller ``screened" problem can be plugged back into the constraints which were neglected, thus offering a quick way to assess if the neglected constraints were actually needed or not. This approach was applied to the UC problem in \cite{Pineda:2020}: based on input loading, a kNN model predicted which lines in the network would \textit{not} be congested. Similarly, \cite{Crozier:2022} used a maximum likelihood formulation to fit Gamma distributions to the probability that thermal and voltage constraints would be active in ACOPF. Future formulations screened out constraints that had a low probability of activation. The procedure accelerated a 24,465-bus ACOPF solve by almost 8-fold. The philosophy behind constraint screening can also be applied to the problem of \textit{contingency} screening, i.e., ignoring the N-1 contingencies, which have little impact on the base case power flow solutions. This was successfully utilized in~\cite{Crozier:2022_pmaps}, where both kNN and NN models were used to screen out unimportant contingencies in networks with up to 31,777 buses. The procedure offered orders of magnitude speedup over conventional screening methods, but the accuracy of the predictions was mixed.

\textit{Using ML for Operational Decision Making}: Other recent advances have focused on training ML and Reinforcement Learning (RL) models to augment, or even replace, the real-time decisions made by human operators in the power grid control room. Sponsored by the French transmission operator (RTE), the Learning 2 Run a Power Network (L2RPN)~\cite{Marot:2020,Marot:2021,Marot:2022} challenge has inspired a host of practical applications in the space of congestion management, topology optimization, and network control. Similar challenges (e.g., ROADEF~\cite{Crognier:2021}) have focused on maintenance prediction and outage planning in large-scale power systems. The Danish TSO Energinet, meanwhile, has partnered with IBM to develop an artificially intelligent ``virtual operator"~\cite{Clemente_2021}. Trained on 400TB of simulation data, this agent will help human operators predict when operational constraints (e.g., N-1 security) could be violated based on historical congestion.

\textit{Training tools as optimizers}: A final promising direction we wish to highlight is related to using ML-inspired training tools (stochastic gradient descent, differentiable layers, Adam, etc.) to solve any number of conventional power system problems, as in DC3~\cite{acopf-dnn-donti}. For example, \cite{Agarwal:2022} used adversarial robustness approaches from the ML literature to pose and solve (via projected gradient descent) a large-scale stochastic optimization problem. Alternatively, \cite{lange2020learning} used ML-inspired strategies to differentiate through a holomorphic load flow solver of a moderately sized system.  Much of this work has been inspired by OptNet~\cite{Amos:2017}, which first proposed the idea of treating an optimization problem as a differentiable \textit{layer} inside of an ML model. New toolboxes from the power systems community, such as Neuromancer~\cite{Neuromancer2022}, allow researchers to more easily extend ML-inspired optimization tools to conventional physics-based problems within the field of power systems.




%


\section{Optimizations for Low/Medium-voltage Distribution Grids} \label{sec:distributed}

Distribution systems before the 2010s were characterized by passive loads, unidirectional flows, and limited/no data beyond the substation SCADA or infrequently-sampled regulator data. The main controllable assets were a small number of front-of-the-meter, tap-changing transformers (at the substation), some capacitor/reactor banks and voltage regulators (sparsely placed throughout feeder), and switches to enable loop scheme operations during contingencies~\cite{paudyal_optimal_2011}. Operations were largely set-it-and-forget-it, open-loop rules (e.g., for cap banks, tap-changers, and regulators) or post-event reactions (e.g., switches after contingency)~\cite{shukla_efficient_2019}. That is, there were little to no dynamic decisions/set-points to optimize, limited data available on which to validate realistic models and, thus, no practical need for optimization. That is also why the IEEE Test Feeder Working Group released a set of test feeders from which optimization schemes could be tested on 3-phase feeders with up to 8500 nodes~\cite{schneider_analytic_2018}.

Fast forward to today's distribution grids. Driven largely by a precipitous drop in the cost of sensors/digitalization and renewable generation, distribution grids are rapidly transforming into data-rich, bidirectional, and active/responsive/dynamic power systems~\cite{almassalkhi_hierarchical_2020,bassi_electrical_2022}. Helped along by advanced metering infrastructure (AMI) and new electrification efforts, medium- and low-voltage grids are experiencing an unprecedented, large-scale penetration of distributed energy resources (DERs), including residential EV chargers, batteries, and smart, but energy-hungry buildings and appliances, each of which can be responsive to incentive and grid signals~\cite{sekhavatmanesh_optimal_2022,almassalkhi_hierarchical_2020,}. 

Today, distribution grids have many decision variables that can benefit from coordination and optimization~\cite{hanif_distribution_2022}. In addition, the data from AMI can support parameter estimation and physical (predictive) model validation, which provides a new paradigm for physics-based optimal power flow in distribution grids. Furthermore, distribution optimal power flow naturally results in larger-scale problems, because all three phases should be considered in unbalanced operations. Additionally, the highly distributed installations of a myriad of DERs, including stochastic resources and bidirectional batteries and EV chargers, begets challenging optimization on the distribution front~\cite{gan_exact_2015}. 

\subsection{Physics-based Solution Techniques}

Physics-based techniques in distribution optimal power flow (DOPF) started with the simplifying but strong assumption that distribution grids were operated as balanced, 3-phase, and radial networks~\cite{baran_optimal_1989,jabr_radial_2006}. This approach gave rise to the so-called non-convex \textit{DistFlow} branch-flow formulation from which numerous researchers have pursued linearizations~\cite{bolognani_existence_2016} and simplifications, such as \textit{LinDistFlow}~\cite{og_linDistFlow}, and various convex relaxations~\cite{farivar_branch_2013, gan_exact_2015}, whose optimal solutions may not be AC-feasible, and convex restrictions~\cite{n_nazir_voltage_2021,n_nazir_grid-aware_2021}, whose solutions are always AC-feasible but can be overly conservative. However, the balanced 3-phase assumption is generally invalid in LV/MV grids (outside of radial wind farm collector networks, which are indeed balanced~\cite{nazir2022:windFarm}). 

Thus, to broaden the appeal of DOPF to more realistic (unbalanced) networks, significant efforts have focused on so-called unbalanced DOPF formulations, which generally employ a non-convex, nonlinear (NLP) DOPF formulation~\cite{paudyal_optimal_2011, bernstein_load_2018, n_nazir_receding-horizon_2018,stai_solution_2021}. Similarly to DistFlow, researchers have approached the unbalanced NLP formulation with various linear approximations (e.g., \textit{Lin3DistFlow}) \cite{arnold_optimal_2016} and convex reformulations~\cite{nick_exact_2018, franco_ac_2018, n_nazir_optimal_2020,jha_network-level_2021}. 

The above methods generally consider a centralized formulation that is sent to a solver (e.g., NLP for non-convex, LP for linear, or SOCP/SDP/QCQP for convex programming solvers) and then returns the optimal solution. NLP formulations are ill-suited for large networks (at the scale of 8500 nodes) and fast timescale DER set-point optimization (e.g., minutely). In addition, even though centralized linear/convex formulations generally scale well, their implementation depends crucially on fully parameterized network models and forecasted/known net loads at individual MV/LV nodes (e.g., representing as little as 1-5 meters). While various stochastic optimization techniques can overcome the effects of forecasts errors and produce (probabilistically) robust optimal solutions~\cite{dallanese_chance-constrained_2017}, the data availability assumptions are rather strong (even today) and limit the applicability and scalability of centralized DOPF approaches~\cite{almassalkhi_hierarchical_2020}. 

Instead, distributed (or non-centralized) implementations have been proposed that effectively update DER set-points to track the (time-varying) optimal solution while only leveraging available (limited) grid data~\cite{dallanese_optimal_2018,bernstein_real-time_2019}. These distributed methods push the DOPF problem into the realm of feedback-based implementations that leverage data streams to correct control set-points to steer the distribution system toward a desirable (optimal) solution while abiding by constraints (e.g., asymptotically, in steady-state). Similar data-driven and machine-learning techniques are being proposed to optimize distribution system operations and represent an interesting path towards scalable DOPF implementations~\cite{chen_icnn_2020,bassi_electrical_2022}.\\

\subsection{Emerging data-driven techniques for distribution networks}
Many of the data-driven and ML-based tools described in the transmission section are also applicable for and have indeed been tested on, distribution networks. For example,~\cite{Manish:2020} used sensitivity-informed NNs to learn inverter dispatch solutions of a low-voltage OPF problem directly. In distribution grids, there is also a strong focus on learning \textit{local} optimization and control policies: to aid in the roll-out of the 1547.8 standard, \cite{gupta2022deep} used NNs to learn optimal and stable volt/VAR injection control policies for distributed energy resources in a distribution grid. The major application of ML tools to distribution grids, however, has been within the realm of estimation and observability, with a recent review paper offered in \cite{Deka:2022}. The general premise is that historical data can be used to infer otherwise-unknown information about distribution grids, e.g., topological information, line parameters, load correlations, etc. Several works in this regard include graphical learning approaches for estimating topologies~\cite{deka2016estimating}, joint topology and parameter estimation~\cite{yu2017patopa}, data-driven model order reduction of distribution system state estimation (DSSE)~\cite{Chevalier:2021}, and Guass-Newton initialization via NNs for robust converge of DSSE~\cite{zamzam2019data}.

\section{Emerging trends in combined T\&D simulations}

The bidirectional interactions between the traditionally \textit{separated} transmission and distribution entities are quickly increasing due to the growing count of resources in the distribution and sub-transmission grid layers. Combined transmission and distribution (T\&D) analysis is touted as one potential solution for evaluating these interactions. Combined T\&D analysis is a large-scale problem integrating features from both high-voltage transmission and low-voltage distribution networks. An average-sized combined T\&D network can have more than a million nodes \cite{TD_synthetic_testcases}, \cite{TD_integrated_pandey} with solution matrix sizes ranging in tens of millions for simulation and optimization problems.

Generally speaking, two broad categories of solution approaches for combined T\&D analyses are emerging: i) the co-simulation simulation (e.g., \cite{TD_cosimulation_FNCS}, \cite{TD_cosimulation_HELICS}, \cite{TD_cosimulation_IGMS}), and ii) the integrated approach (e.g., \cite{TD_integrated1}, \cite{TD_integrated_pandey}, \cite{TD_integrated3}). Co-simulation approach uses disparate tools for transmission and distribution simulations and supports communication between the tools to reach a consensus. In the integrated approach, a unified framework solves both T\&D networks using a singular algorithm and internally communicates between distributed compute resources to solve the overall network. A good review of these two viewpoints for combined T\&D modeling approaches can be found in \cite{TD_jain_survey}. 

Specifically, on the combined T\&D optimization front, we find that the literature is limited, and the field is still emerging. U.S. national labs have ongoing projects exploring the joint T\&D ACOPF problem with open-source models \cite{TD_opti_national_labs}. There is also recent literature on optimal control of joint transmission and distribution grids; however, empirical results are mostly for smaller-sized networks \cite{TD_opti_control}. Another recent research uses combined T\&D co-optimizations to dispatch DER resources \cite{ED_TD_NREL}. Finally, there are also preliminary efforts on using optimization to estimate states of combined T\&D networks \cite{se_dist_global}, \cite{TD_SE}.

The big limitation for research on combined T\&D optimization is the lack of large-scale realistic models and standardization of input data that various approaches can ingest. Generally, validating, benchmarking, and testing different combined T\&D approaches require synthetic but realistic large-scale test networks. On the simulation front, we are beginning to observe \textit{much-valued} advancements (see some examples in \cite{TD_synthetic_testcases}, \cite{TD_testcases}), but such input models and data remain limited on the optimization front.  \label{sec:combined}

\section{Observable Gaps in Large-scale Optimization} \label{sec:gaps}

In systemizing the state-of-the-art in large-scale grid optimization, we sought feedback from industry partners on \textit{observable gaps} in the field. We shared the manuscript or highlighted the key analyses covered within it to facilitate feedback. At a high level (from discussions in \cite{isone_private}, \cite{kwok_private}, and \cite{marko_private}), we found several disconnects between industry needs and current academic research in the field of large-scale optimization. Still, we also observed increased collaboration between industry and academia, primarily because of large, funded programs on the subject from various federal entities (e.g., ARPA-E in the U.S. and RTE in France) In general, industry experts highlighted several challenges and gaps in large-scale grid optimization. \cite{marko_private} noted that the input model quality is a big challenge and that there is a need for optimization methods that are robust to erroneous models. \cite{marko_private} further stated that many industry processes today solve an optimization through brute force by running 1000s of simulations and that new optimization methods can replace some of these processes via single optimization run. \cite{isone_private} and \cite{kwok_private} called for improved market algorithms for a larger footprint, with more resources (e.g., DERs, VPPs, co-located resources), and for a finer time resolution and over a longer time horizon. They also desired to include these features without incurring any overhead on the run-time. \cite{kwok_private} emphasized the need for methods for distribution market optimization and multi-level optimization coordinating between ISO markets and DSO markets. \cite{kwok_private} also cited the need for more accurate models for managing features of new resources, such as the state-of-the-charge of storage and  transitioning between various combined cycle configurations. \cite{isone_private} noted the future need for stochastic analyses such as stochastic unit commitment, especially when the system is constrained by fuel supply, as is New England in winter months. On the distribution front, the lack of models was discussed as the biggest challenge. But, we find that with increasing frequency and locations of data capture and transmittal  \cite{cyril_private}, model quality will likely improve; and can act as a catalyst for field deployments of newer, more practical methods for DER control and dispatch. 

\section*{Annotated References}

\begin{enumerate}
    \item $\bullet$ Y. Chen, F. Pan, J. Holzer, E. Rothberg, Y. Ma and A. Veeramany, ``A High Performance Computing Based Market Economics Driven Neighborhood Search and Polishing Algorithm for Security Constrained Unit Commitment," in IEEE Transactions on Power Systems, vol. 36, no. 1, pp. 292-302, Jan. 2021, doi: 10.1109/TPWRS.2020.3005407.
    \underline{Annotations:} This paper is from an ARPA-E awarded HIPPO project that built parallel and distributed computing capabilities for real-world SCUC market algorithms. This particular paper develops a neighborhood search and polishing algorithm that adaptively fixes binary and continuous variables and chooses lazy constraints based on hints from an initial solution and its associated neighborhood.

    \item $\bullet$ Park, S., Chen, W., Mak, T. W., \& Van Hentenryck, P. (2023). Compact Optimization Learning for AC Optimal Power Flow. arXiv:2301.08840. 
    \underline{Annotations:} This paper uses advanced statistical compression combined with neural network training to build a model which predicts ACOPF solutions to power system test cases with up to 30,000 buses. This is one of the largest known instances of end-to-end machine learning being used to predict ACOPF solutions, and it exemplifies how data-driven and learning-based innovations can help aid and accelerate large-scale grid computations.
    
    \item $\bullet$ Tsai, C. H., Figueroa-Acevedo, A., Boese, M., Li, Y., Mohan, N., Okullo, J., ... \& Bakke, J. (2020). Challenges of planning for high renewable futures: Experience in the US midcontinent electricity market. Renewable and Sustainable Energy Reviews, 131, 109992.
    \underline{Annotations:} MISO developed the Renewable Integration Impact Assessment (RIIA) framework to study high renewable scenarios (up to 50\%) in their footprint. To assess system impact in this study, MISO ran a suite of large-scale grid optimizations, ranging from production cost modeling to ACOPF.
\end{enumerate}

\begin{acknowledgements}
The authors like to thank industry members who helped shape the direction of this paper and provided expert feedback on observable gaps in the field of large-scale optimization from the industry's viewpoint. These include Kwok Cheung from GE Grid Solutions, Xiaochuan Luo and Jinye Zhao from ISO New England, Dan Kopin from VELCO, Cyril Brunner from Vermont Electric Cooperative, and Marko Jereminov from Pearl Street Technologies, Inc. In addition, M. Almassalkhi gratefully acknowledges support from National Science Foundation awards ECCS-2047306.
\end{acknowledgements}


\section*{Statement and Declarations}

\subsection*{\textbf{Funding}}

M. Almassalkhi gratefully acknowledges support from National Science Foundation awards ECCS-2047306.

\subsection*{\textbf{Competing Interests}}

A. Pandey owns equity and consults for clean-tech startup Pearl Street Technologies, Inc.

\subsection*{\textbf{Author Contributions}}

All authors contributed to the study conception and design. A.P led the work on mechanistic physics-based methods for transmission and combined T\&D networks, S.C led the work on data-driven methods for both T and D networks, and M.A led the work on physics-based methods for distribution grids. All authors read and approved the final manuscript.

\subsection*{\textbf{Ethics Approval and Consent to Participate}}
Not-applicable

\printbibliography
\end{document}